\documentclass[a4paper]{article} 

\usepackage{endfloat}
\usepackage{latexsym}
\usepackage{amsmath}
\usepackage{graphicx}
\usepackage{amssymb}

\newcommand\comment[1]{} 
\newcommand{\sizedef}{
      \headheight=0pt                               
	  \topmargin=-1.5cm \headsep=1.5cm              
      \oddsidemargin=-0.5cm \evensidemargin=-0.5cm  
      \textheight=22truecm \textwidth=16.5truecm    
}
\newtoks\reportnoregister \newtoks\eprintnoregister
\newcommand{\reportnumber}[1]{\reportnoregister={#1}}
\newcommand{\eprintnumber}[1]{\eprintnoregister={#1}}

\reportnumber{\mbox{}} 
\eprintnumber{\mbox{}} 

\newcommand{\reportid}{
   \begin{minipage}{17cm}\vspace{-3.2cm}
     \begin{flushright}
      {\normalsize \the\reportnoregister \\[-.2cm]
	    \eprint{\the\eprintnoregister}}\vspace{3.2cm}
     \end{flushright}
   \end{minipage}\hspace{-17cm} }

\catcode`@=11   
\def\title#1{\gdef\@title{\reportid#1}}
\catcode`@=12   

\newcommand{\eprint}{\textsf} 

\newcommand{\journalfont}{\rm}  
\newcommand{\jou}[1]{{\journalfont #1\ }}
\newcommand{\joudef}[2]{\newcommand #1{\jou{\ignorespaces #2}}}

\joudef{\aaa}    { Astron.\ Astrophys.}
\joudef{\aip}    { Adv.\ Phys.}
\joudef{\adm}    { Adv.\ Math.}
\joudef{\am}     { Ann.\ Math.}
\joudef{\apny}   { Ann.\ Phys.\ (N.Y.)}
\joudef{\apj}    { Astrophys.\ J.}
\joudef{\apjs}   { Astrophys.\ J.\ Suppl.}
\joudef{\cjp}    { Can.\ J.\ Phys.}
\joudef{\cmp}    { Commun.\ Math.\ Phys.}
\joudef{\cqg}    { Class.\ Quantum Grav.}
\joudef{\faa}    { Funct.\ Anal.\ Appl.}
\joudef{\grg}    { Gen.\ Rel.\ Grav.}
\joudef{\ijmpd}  { Int.\ J.\ Mod.\ Phys.\ D}
\joudef{\ijtp}   { Int.\ J.\ Theor.\ Phys.}
\joudef{\invm}   { Invent.\ Math.}
\joudef{\jm}     { J.\ Math.}
\joudef{\jmp}    { J.\ Math.\ Phys.}
\joudef{\jpa}    { J.\ Phys.\ A}
\joudef{\mnras}  { Mon.\ Not.\ R.\ Ast.\ Soc.}
\joudef{\mpla}   { Mod.\ Phys.\ Lett.\ A} 
\joudef{\nature} { Nature}
\joudef{\nc}     { Nuovo Cim.}
\joudef{\npb}    { Nuc.\ Phys.\ B}
\joudef{\ph}     { Physica}
\joudef{\pla}    { Phys.\ Lett. A}
\joudef{\plb}    { Phys.\ Lett. B}
\joudef{\pr}     { Phys.\ Rev.}
\joudef{\prd}    { Phys.\ Rev.\ D}
\joudef{\prep}   { Phys.\ Rep.}
\joudef{\prl}    { Phys.\ Rev.\ Lett.}
\joudef{\prsla}  { Proc.\ Roy.\ Soc.\ Lond.\ A}
\joudef{\ptp}    { Prog.\ Theor.\ Phys.}
\joudef{\ptps}   { Prog.\ Theor.\ Phys.\ Suppl.}
\joudef\rmp      { Rev.\ Mod.\ Phys.}
\joudef\spj      { Sov.\ Phys.\ JETP}

\catcode`@=11

\newcommand\eqalign[1]{\null\,\vcenter{\openup\jot\m@th
  \ialign{\strut\hfil$\displaystyle{##}$&$\displaystyle{{}##}$\hfil
      \crcr#1\crcr}}\,}
\newcommand\meqalign[1]{\null\,\vcenter{\openup\jot\m@th
  \ialign{\strut\hfil$\displaystyle{##}$&&$\displaystyle{{}##}$\hfil
      \crcr#1\crcr}}\,}
\def\ps@reportnumber{%
    \let\@oddfoot\@empty\let\@evenfoot\@empty
    \def\@oddhead{\hfil\rightmark}}
	
\catcode`@=12   

\newdimen\arrayruleHwidth
\makeatletter
\newcommand\Hline{\noalign{\ifnum0=`}\fi\hrule \@height \arrayruleHwidth
  \futurelet \@tempa\@xhline}
\makeatother

\newcommand\thickbaselines{\baselineskip=20pt\lineskip=3pt\lineskiplimit=3pt}

\catcode`@=11

\renewcommand\matrix[1]{\null\,\vcenter{\thickbaselines\m@th
    \ialign{\hfil$##$\hfil&&\quad\hfil$##$\hfil\crcr
      \mathstrut\crcr\noalign{\kern-\baselineskip}
      #1\crcr\mathstrut\crcr\noalign{\kern-\baselineskip}}}\,} 
\catcode`@=12   

\newcommand\be{\begin{equation}} \newcommand\ee{\end{equation}} 
\newcommand\bd{\begin{displaymath}}\newcommand\ed{\end{displaymath}}

\renewcommand{\d}{{\rm d}} 

\newcommand\undersim[1]{\mathop{\vtop{\ialign{##\crcr
     $\hfil\displaystyle{#1}\hfil$\crcr\noalign
     {\kern1pt\nointerlineskip}\hbox{$\hfil\sim\hfil$}\crcr
     \noalign{\kern1pt}}}}}

\newcommand\case[2]{\textstyle{\frac{#1}{#2}}}
\newcommand\dsfrac[2]{\displaystyle{\frac{#1}{#2}}}

\newcommand{\smallcaption}[1]{\caption{\protect\small#1}}

\newcommand{\acronym}[3]{\newcommand{#1}{#3 (#2)\relax\renewcommand{#1}{#2}}}

\sizedef

\acronym{\hw}{HW}{{Harrison-Wheeler}}
\acronym{\gbone}{GB1}{{\em generalized Buchdahl $n=1$ polytrope}}
\acronym{\gbfive}{GB5}{{\em generalized Buchdahl $n=5$ polytrope}}
\acronym{\sss}{SSS}{{\em static spherically symmetric}}

\begin{document}

\reportnumber{USITP 98-16}
\eprintnumber{gr-qc/9809033}

\title{\Large Trapped gravitational wave modes in stars with $R>3M$.}
	   \author{Kjell Rosquist \\
        {\small Department of Physics, Stockholm University}  \\[-10pt]
        {\small Box 6730, 113 85 Stockholm, Sweden} \\
        {\small E-mail: \textsf{kr@physto.se}}}
\date{}

\maketitle

\begin{abstract}
The possibility of trapped modes of gravitational waves appearing in stars 
with $R>3M$ is considered.  It is shown that the restriction to $R<3M$ in 
previous studies of trapped modes, using uniform density models, is not 
essential.  Scattering potentials are computed for another family of analytic 
stellar models showing the appearance of a deep potential well for one model 
with $R>3M$.  However, the provided example, although having a more realistic 
equation of state in the sense that $v_{\rm sound} < \infty$, is unstable.  On 
the other hand it is also shown that for some stable models belonging to the 
same family but having $R<3M$, the well is significantly deeper than that of 
the uniform density stars.  Whether there are physically realistic equations 
of state which allow stable configurations with trapped modes therefore 
remains an open problem.

\end{abstract}


\section{Introduction}

Examples of trapped modes of gravitational waves in compact stars were first 
given by Chandrasekhar and Ferrari \cite{cf:gwresonance} and calculations were 
subsequently also carried out by other authors 
\cite{kokkotas:axialmodes,akk:oscillation,kak:oscillation}.  The fundamental 
reason behind the occurrence of the trapped gravity wave modes is the 
stretching of the geometry by the strong gravitational field leading to a 
bell-like geometrical structure inside the star.  This phenomenon is most 
clearly illustrated using the concept of the optical geometry as developed by 
Abramowicz and coworkers (see \cite{aabgs:optical} and references therein).  
The optical geometry of the vacuum Schwarzschild metric develops a neck 
precisely at $R=3M$ implying that for stars with $R<3M$ there will be a family 
of closed null geodesics in the stellar interior.  It is natural to associate 
this behavior with the trapping of certain modes of gravitational radiation 
although the relation between the trapping and appearance of the neck in the 
optical geometry is only approximate.  The optical geometry is useful not only 
for pedagogical purposes but can also be used to motivate an estimate of the 
eigenfrequencies of the resonances \cite{aabgs:optical}.  In previous studies 
it has often been assumed that the appearance of a neck (and the consequent 
trapping of gravity waves) is only possible if the star is ultracompact, that 
is the {\em compactness\/}, $\beta := M/R$, must lie in the range $\case13 < 
\beta < \case49$ where the upper bound $\case49$ is Buchdahl's limit 
\cite{buchdahl:limit} representing the maximum compactness for any static star 
for which the energy density is decrasing outwards.  The compactness is 
usually given in terms of the inverse compactness $\alpha := R/M$ which we 
will refer to as the {\em tenuity\/}.  The trapped modes found in 
\cite{cf:gwresonance} occur for tenuities in the range $2.25 < \alpha \lesssim 
3$.  Realistic neutron stars are believed to have tenuities in the range $3 
\lesssim \alpha \lesssim 11$ so they are at most marginally ultracompact in 
this sense \cite{ivd:ultracompact}.  However, as will be shown in this letter, 
trapped modes may occur in stars with $\alpha >3$.  This opens up the 
possibility for real neutron stars to exhibit gravity wave trapping.  In view 
of this result it seems like a good idea to reserve the notion of 
ultracompactness for stars which have a neck in their optical geometry and 
consequently a family of closed null geodesics in their interior.  
Ultracompact stars would then be expected to exhibit gravity wave resonances 
as well.  As will become more clear later ultracompactness in this sense 
really applies to the stellar core rather than the entire star.  Although the 
definition of ultracompactness given here is unambiguous it is more difficult 
to calculate in practice.  In the concluding remarks we will touch upon 
possible rules of thumb criteria which could be used as a rough estimate of 
compactness.

It is not difficult to understand why stars with $R>3M$ could have an 
ultracompact core.  The key is the behavior of the equation of state at low 
pressure.  Consider a uniform density model with radius less than $3M$.  Now 
replace a thin shell (its mass should be finite but be only a small fraction 
of the total mass) at the surface with some material with a soft equation 
state (for example a polytrope) such that the total mass of the star remains 
the same.  In physical terms we can think of this process as giving the star 
an atmosphere by transforming some of the matter near its surface.  Clearly 
the gravitational field in the core is the same as it was before.  However, 
the radius will depend sensitively on the equation of state of the atmosphere.  
In fact it can be made arbitrarily large for example by letting the atmosphere 
be a polytrope of index $5-\epsilon$ where $\epsilon \ll 1$.  Another 
alternative would be to replace the slice by an envelope which, like the core 
matter, is of uniform density but satisfying $\rho_{\rm envelope} < \rho_{\rm 
core}$.  Such double layer uniform density models were recently considered by 
Lindblom \cite{lindblom:phase} to discuss phase transitions in compact stellar 
models.  The radius could then be made aribtrarily large by letting the 
quotient $\rho_{\rm envelope}/\rho_{\rm core}$ be sufficiently small.

Although the argument given above should be sufficient to establish the 
existence of trapped gravity wave modes for stellar models with $R>3M$, there 
remains some critical issues concerning the realization of such models in 
nature.  One such issue is the question of causality.  Of course, already the 
unform density models are unrealistic in this sense having an infinite speed 
of sound.  A second issue is that of stability.  The absence of a local mass 
maximum in the uniform density models shows that they are in fact stable.  In 
this letter we will use the \gbfive\ family of exact models 
\cite{simon:polyfive,rosquist:star} to illustrate the new possibilities which 
occur when one considers softer equations of state.  This family generalizes 
the original Buchdahl solution which behaves as a polytrope of index 5 at low 
pressure.  The generalized models, however, have an equation of state which is 
liquid-like at low pressure in the sense of having $\rho_{\rm s} >0$ (``s'' 
denoting the value at the stellar surface).

\section{Stellar models}

The metric of \sss\ models is usually given in the Schwarzschild form
\begin{equation}\label{eq:metric1}
	g = -e^{2\nu}\d t^2 + e^{2\lambda}\d r^2
	                       + r^2(\d\theta^2 + \sin^2\!\theta\,\d\phi^2) \ .
\end{equation}
For our purposes we also need to write the metric of a \sss\ system in a 
general radial gauge as
\begin{equation}\label{eq:metric2}
	g = -Y^2 \d t^2 + N^2 \d x^2
	                       + S^2 (\d\theta^2 + \sin^2\!\theta\,\d\phi^2) \ ,
\end{equation}
where $Y$, $N$ and $S$ are functions of the radial variable $x$.  The 
Schwarzschild radial variable is then given by the relation $r=S(x)$.
Before proceeding we need to deal with a possible source of confusion relating 
to the metrics \eqref{eq:metric1} and \eqref{eq:metric2}.  The time coordinate 
is {\em a priori\/} only defined up to a scaling and a translation.  The 
scaling gauge can be fixed by the requirement that the time coordinate should 
correspond to the proper time of a static observer at infinity.  We shall 
refer to this gauge as the proper time gauge.  This gauge is usually but not 
always imposed when writing down the metric of exact solutions.  It is assumed 
here that the metrics \eqref{eq:metric1} and \eqref{eq:metric2} refer to the 
proper time gauge.  Correspondingly the formulas given below are also given in 
this gauge.  However, since exact solutions are not automatically given in 
the proper time gauge it is useful to write down the relevant transformation 
formula for a metric written in a general time gauge.
To do that we first note that for the Schwarzshild exterior metric (as usual 
expressed in the proper time gauge) $k:= \sqrt{1-2M/R} = e^{\nu_{\rm s}}
=Y_{\rm s}$, where the  subscript $s$ denotes the surface of the star.  
Therefore we must have $e^{\nu_{\rm s}} = k$ for the stellar model.  Now let
$\tilde t$ be an  arbitrary time coordinate and $\tilde\nu$ (or $\tilde Y$)
the corresponding  metric functions.  Then the required relations are
\begin{equation}
	e^\nu = k e^{\tilde\nu-\tilde\nu_{\rm s}} \ ,\qquad
	    Y = k \tilde Y_{\rm s}^{-1} \tilde Y \ ,\qquad
		\d\tilde t = k \tilde Y_{\rm s}^{-1} \d t \ .
\end{equation}

The gravity wave modes discussed in \cite{cf:gwresonance} are equivalent to 
non-radial axial ({\em i.e.\ } odd parity) perturbation modes of \sss\ fields.  
Such axial perturbations do not couple to fluid motions in the star a fact 
which accounts for their alternative interpretation as gravitational wave 
modes.  The axial modes with frequency $\omega$ and mode number $l \geq 2$ are 
governed by the equation \cite{cf:osc}
\begin{equation}
	-\frac{\d^2 Z}{\d {r_*}\!\!^2} + V Z = \omega^2 Z \ ,
\end{equation}
where the potential $V = V_l + V_{\rm d}$ is here written formally as a sum of 
a centrifugal and a dynamical part ({\em cf.} \cite{sm:opticgeom}) in the form
\begin{equation}
 \begin{split}
          V_l &= l(l+1)e^{2\nu}r^{-2} \ ,\\[5pt]
	V_{\rm d} &= e^{2\nu}\left[ 
		      r^{-1}\left(\lambda_{,r}-\nu_{,r}+2r^{-1}\right)e^{-2\lambda}
		          -2r^{-2} \right] \ ,
 \end{split}
\end{equation}
and $r_*$ is the tortoise radial variable defined by
\begin{equation}
	\d r_* = e^{\lambda-\nu}\d r = Y^{-1}N \d x \ .
\end{equation}
To express the potential in a general radial gauge we use the relations 
$e^\nu = Y$, $e^\lambda = N{S'}^{-1}$ and
\begin{equation}
 \begin{split}
	\lambda_{,r} &= N^{-1}N'\,{S'}^{-1} - {S'}^{-2}S'' \ ,\\
    	\nu_{,r} &= Y^{-1}\,Y'\,{S'}^{-1} \ ,
 \end{split}
\end{equation}
where the primes denote differentiation with respect to $x$. The potentials 
then become
\begin{equation}
 \begin{split}
	   V_l &= l(l+1)Y^2S^{-2} \ ,\\
	V_{\rm d} &= Y^2N^{-2}S^{-2}\left[\left( N^{-1}N'-Y^{-1}Y' \right)SS'
	              - SS'' + 2{S'}^2 - 2N^2 \right] \ .
 \end{split}
\end{equation}
Using the Einstein equations the dynamical part of the potential can be 
written in the form
\begin{equation}
	V_{\rm d} = G e^{2\nu}\left[
	                         4\pi(\rho - p) - \frac{6 m(r)}{r^3}\right] \ ,
\end{equation}
where
\begin{equation}
	m(r) = 4\pi\int_0^r \rho r^2 \d r 
	     = 4\pi\int_{x_{\rm c}}^{x} \rho S^2 S' \d x \ ,
\end{equation}
is the mass within radius $r$.  We are using units in which $c=1$ but keep the 
gravitational constant, $G$, for convenience in some formulas.  Geometric 
units can be obtained by setting $G=1$.
%

\subsubsection*{The exterior Schwarzschild solution}
In this case
\begin{equation}
	e^{2\nu} = 1-\frac{2GM}{r} \ ,
\end{equation}
and $m(r)=M$ leading to
\begin{equation}
 \begin{minipage}{12cm}
	$$V_l = \frac{l(l+1)(1-2GM/r)}{r^2} \ ,\qquad
	V_{\rm d} = -\frac{6GM(1-2GM/r)}{r^3} \ ,$$
\vspace{3pt}
	$$ V = \frac{(1-2GMr)[l(l+1)-6GM/r]}{r^{2}} \ .$$
 \end{minipage}
\end{equation}

\subsubsection*{The interior Schwarzschild solution}
Schwarzschild's uniform density model is characterized by
\begin{equation}
  \begin{split}
	e^\nu &= \case12 (3k-y) \ ,\qquad e^\lambda = y^{-1} \ ,\\[5pt]
	y &:= \sqrt{1-\frac{2G\beta^3r^2}{M^2}} \ ,\qquad
	\rho = \frac{3\beta^3}{4\pi M^2} \ ,\qquad
	k = \sqrt{1-2G\beta} \ ,
  \end{split}
\end{equation}
where $\beta:=M/R$.  The  potentials then become
\begin{equation}
	V_l = \frac{l(l+1)(3k-y)^2}{4r^2} \ ,\qquad 
	V_{\rm d} = -4\pi G k\rho \,e^\nu
	          = -\dsfrac{3G \beta^3 \,k(3k-y)}{2 M^2} \ .
\end{equation}

\subsubsection*{The GB5 interior solutions}
The family of exact interior solutions which we focus on this letter is the 
\gbfive\ family \cite{simon:polyfive,rosquist:star} given by (using a 
non-proper time gauge)
\begin{equation}
	\tilde Y = \frac{T-X}{T+X} \ ,\qquad N = S = (T+X)^2 \ ,
\end{equation}
where
\begin{equation}
	T(x) = \sqrt{\frac{\lambda^{-3}\,b}{\cosh(x-\Delta)}} \ ,\qquad
	X(x) = \sqrt{\frac{b}{\cosh x}} \ .
\end{equation}
The constants $b$ and $\lambda$ characterize the equation of state while 
$\Delta$ is the single nontrivial integration constant appearing in all \sss\ 
models.  The equation of state can be written in the form
\begin{equation}\label{eq:eqstate}
      p = \frac{a(u^6-\lambda^6)}{(1+u)^5 (1-u)} \ ,\qquad
   \rho = \frac{3a(u^5+\lambda^6)}{(1+u)^5} \ ,
\end{equation}
where $a := 1/(8\pi G \,b^2)$, $u:= X/T$ and $\lambda := u_{\rm s}$.  Defining 
$\chi := u_{\rm c} = \lambda^{3/2} e^{\Delta/2}$ (``c'' denoting value at the 
center) we have
\begin{equation}
   0 < \lambda = u_{\rm s} < u < u_{\rm c} = \chi < 1 \ . 
\end{equation}
The two parameters $a$ and $\lambda$ which characterize the equation of state 
in \eqref{eq:eqstate} can be interpreted as a scaling and a stiffness 
parameter respectively.  The scaling parameter $a$ just represents a change of 
overall scale.  All other physical characteristics in the model are unaffected 
by changes in $a$ which can be any positive number.  It is convenient to 
replace $\chi$ by another parameter $\mu$ defined by the relation $\chi= 
\lambda+\mu -\lambda \mu$.  In that way the $\lambda\mu$-section of the 
parameter space is exactly the unit square, $0<\lambda<1$, $0<\mu<1$ (see 
figure 1).  However, in order not to complicate the formulas unnecessarily we 
keep using $\chi$ but think of it as a function of $\lambda$ and $\mu$.  For 
our purposes it is also useful to replace $a$ by the mass of the star.  
Expressing $a$ (or $b$) and $\Delta$ in terms of $M$, $\lambda$ and $\chi$ we 
have
\begin{equation}
	T(x) = \sqrt{\frac{2\,\chi^2 \,b}{\lambda^6 e^x + \chi^4 e^{-x}}} \ ,
\end{equation}
where
\begin{equation}
	b = \frac{(\chi^2+\lambda^3)(\chi^2-\lambda^3)^3 M}
	        {4\chi(\chi^2-\lambda^2)^{3/2}(\chi^2-\lambda^4)^{3/2}} \ .
\end{equation}
In calculations we wish to use the set $(\lambda, \alpha, M)$ as input 
parameters to specify the stellar model.  The tenuity is given by the 
expression
\begin{equation}
	\alpha = \frac{(1+\lambda)^2(\chi^2-\lambda^3)^2}
	              {2\lambda(\chi^2-\lambda^2)(\chi^2-\lambda^4)} \ .
\end{equation}
In order to use $\alpha$ as an input parameter we solve this equation for 
$\chi$ which yields
\begin{equation}
	\chi = \lambda^{3/2} \left( \frac{1+\zeta}{1-\zeta}
	                                                \right)^{1/4} \ ,\qquad
	\zeta := \frac{(1-\lambda^2)\sqrt{\alpha(\alpha-2)}}
	          {(1+\lambda^2)\alpha-(1+\lambda)^2} \ .
\end{equation}
In the radial gauge $N=S$ the expression for $V_{\rm d}$ reduces to
\begin{equation}
	V_{\rm d} = Y^{-2}S^{-4}\bigl(3{S'}^2 - Y^{-1}Y'SS' - SS''
	                                                   - 2S^2 \bigr) \ .
\end{equation}
Inserting the \gbfive\ functions $Y$ and $S$ in this expression gives the 
potential in an explicit but complicated form and we do not write it down 
here.

\section{Discussion}

We now consider the new possibilities which occur when using an equation of 
state with non-uniform density using the \gbfive\ family as a theoretical 
laboratory.  The parameter space of the \gbfive\ models is shown in 
figure~1.
\begin{figure}[!tbp]
\centering
\includegraphics[width=10cm]{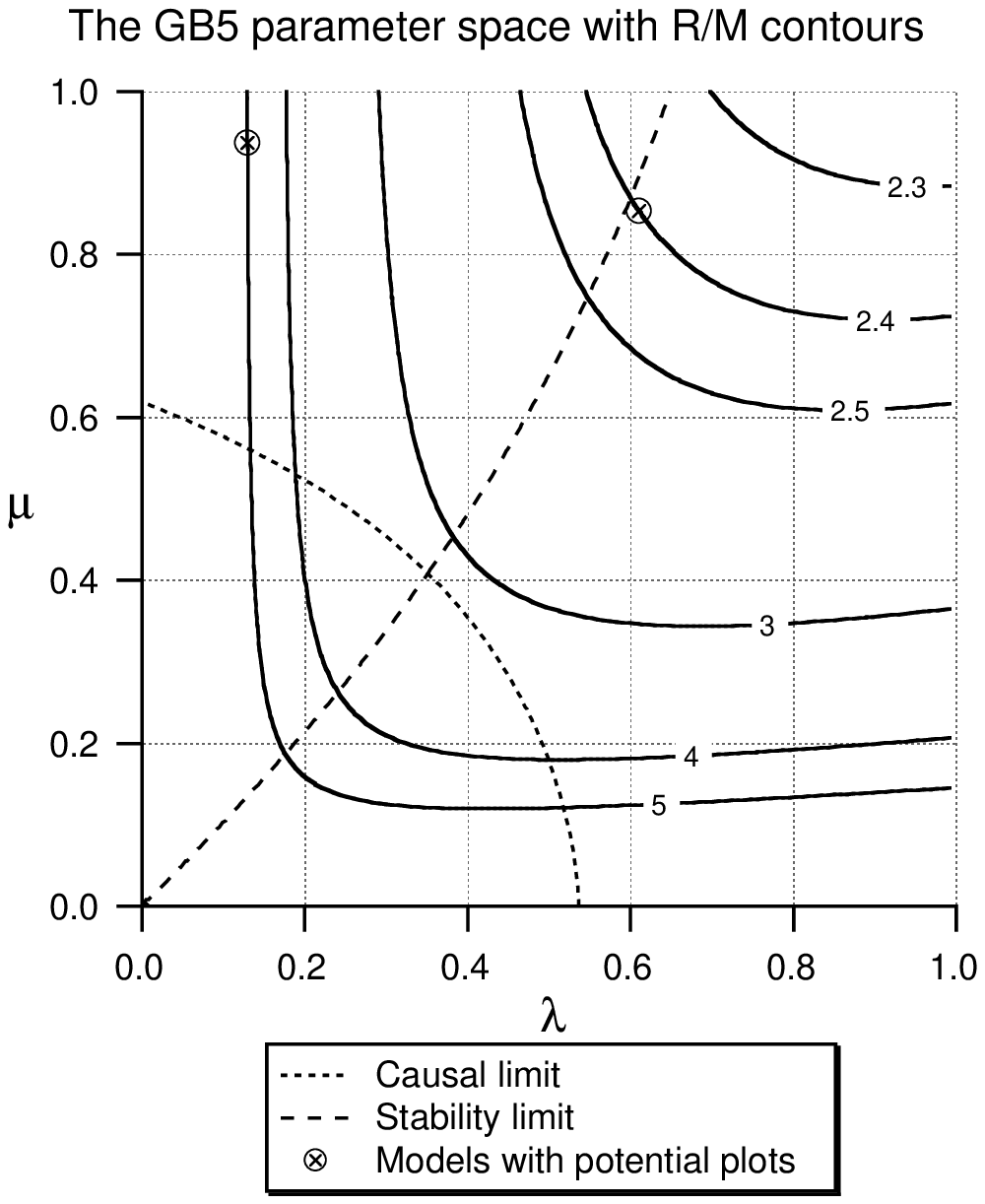}
\begin{minipage}{12cm}
\smallcaption{ The $\lambda\mu$ parameter space of the \gbfive\ family.  The 
pressure and energy density satisfy the physical requirements $p\geq0$, 
$\rho\geq0$ and $\d p/\d\rho \geq0$.  Contours are shown for some values of 
the tenuity $\alpha =R/M$.  Specifying the value of $\lambda$ fixes the 
equation of state up to a scaling.  The corresponding sequence of stellar 
models can then be followed along the $\lambda= constant$ line starting from 
the Newtonian limit at $\mu=0$.  The mass increases along such a sequence up 
to the dashed curve after which it decreases.  The dashed curve therefore 
represents the transition to unstable models.  The stable region is 
consequently located to the lower right of the stability limit.  The dotted 
curve is the causal limit.  A model is classified as causal if the equation of 
state satisfies $\d p/\d\rho \leq1$ throughout the interior of the star.  For 
acausal models the speed of sound as defined by $v_{\rm sound} := \sqrt{\d 
p/\d\rho}$ is therefore larger than the speed of light in some part of the 
stellar interior.  The causal \gbfive\ systems are located at the lower left 
side of the dotted curve in parameter space.  The two marks in the figure are 
the points which correspond to the models for which potential plots are given 
in this letter.  }\end{minipage}
\label{fig:GB5}
\end{figure}
One of the two models marked in figure~1 has $\alpha = 5$.  The scattering 
potential for that model is shown in figure~2.
\begin{figure}[!tbp]
\centering
\hspace{-2cm}\includegraphics[width=16cm]{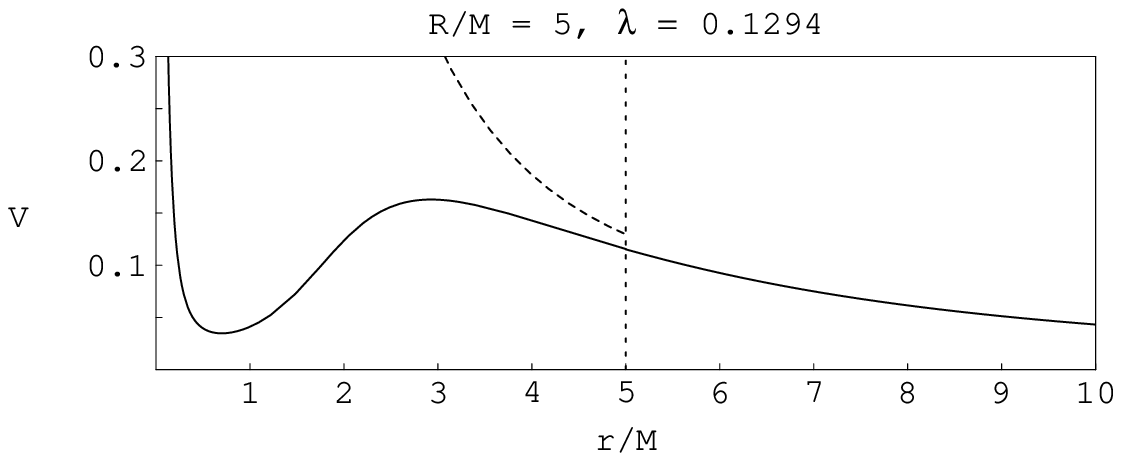}
\begin{minipage}{12cm}
\smallcaption{The \gbfive\ potential for the model marked on the $\alpha=5$ 
contour in figure~1 is plotted for $l=2$ along with the exterior potential.  
For comparison the potential of the uniform density model with the same mass 
and radius is shown as the dashed curve.  }
\end{minipage}
\label{fig:potential5_1294}
\end{figure}
This clearly illustrates the fact that the potential may have a minimum in the 
stellar interior even though the outer parts of the star extend to regions 
well beyond $\alpha = 3$.  We also mention without proof that this model 
admits a family of closed null geodesics in its interior.  It may be objected 
that the model is unstable (as indicated in figure~1) and that this result 
therefore has little physical relevance.  However, the instability is closely 
connected with the softness of the equation of state.  Taking instead the 
double layer uniform density models mentioned in the introduction it should be 
possible to provide examples of stable models having a potential with a 
minimum in the interior.
A second comment we wish to make on this issue is that resonances in unstable 
models may in principle be important in gravitational collapse situations 
where short-lived unstable equilibrium states could perhaps form en route to 
the final collapse.

The second model indicated in figure~1 has $\alpha=2.4$ and lies in the stable 
region of the parameter space.  The corresponding potential is plotted in 
figure~3.
\begin{figure}[!tbp]
\centering
\hspace{-2cm}\includegraphics[width=16cm]{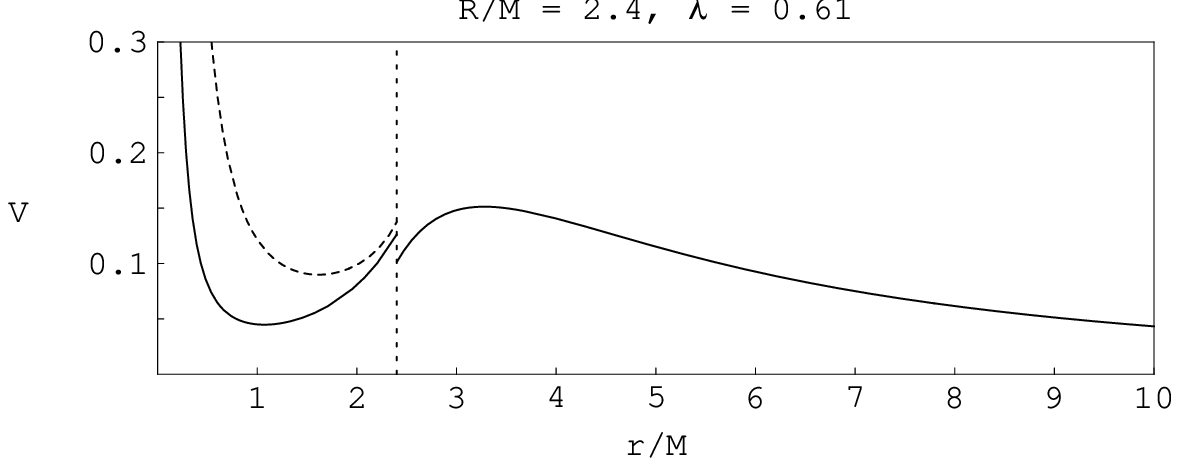}
\begin{minipage}{12cm}
\smallcaption{The potential for the second \gbfive\ model marked in figure~1 
is plotted for $l=2$.  As in figure~2 the dashed curve is the potential of the 
corresponding uniform density model.  }\end{minipage}
\label{fig:potential24_61}
\end{figure}
The phenomenon we wish to illustrate here is that the \gbfive\ potential has a 
significantly deeper potential well than a uniform density model with the same 
mass and $R/M$ ratio.  The quasi-normal modes of the uniform density model 
with $\alpha =2.4$ were calculated in \cite{cf:gwresonance}.  It would be 
interesting to calculate the modes for the \gbfive\ model.  The deeper minimum 
is an indication of longer damping times compared to the uniform density case.

The question of whether realistic stellar models can be ultracompact (in the 
sense defined in this letter) remains open.  In \cite{ivd:ultracompact}, Iyer, 
Vishveshwara and Dhurandar searched for stable and causal models satisfying 
$R<3M$.  In view of the results given in the present work it would be more 
relevant to look for stable and causal models which are ultracompact in the 
sense of having a family of closed null geodesics in the stellar interior.  It 
would be very useful to have a simple rough criterion of ultracompactness 
expressed in terms of a dimensionless combination of easily computable 
quantities.  Examples of criteria of compactness include the central redshift 
(defined as the redshift of a hypothetical speed of light signal sent from the 
center of the star and received by a static observer at infinity) and the 
central 4-dimensional curvature, for example $(R^{\alpha\beta\gamma\delta} 
R_{\alpha\beta\gamma\delta})_{\rm c}$ or $(R^{\alpha\beta} 
R_{\alpha\beta})_{\rm c}$.  The redshift is already dimensionless while the 
curvature measures need to be properly normalized for example by multiplying 
by a power of the total mass.  However, it is not clear whether any of these 
measures, either by themselves or by taking combinations, could serve as 
criteria for ultracompactness.

\section*{Acknowledgements}

This work could not have been carried out without the influence of a number of 
colleagues.  I would like to thank in particular Marek Abramowicz,
Joachim Almergren, Ingemar  Bengtsson, Emanuele Berti, Valeria Ferrari,
S\"oren Holst and Remo Ruffini for  their interest and valuable comments. 
Many thanks are also due to Giuseppe  Pucacco and the ICRA group at the
University of Rome for providing the  stimulating environment where this work
was completed.  Financial support was  given by the Swedish Natural Science
Research Council.


\end{document}